\documentclass[conference]{IEEEtran}
\IEEEoverridecommandlockouts

\usepackage{cite}
\usepackage{amsmath, amssymb}
\usepackage{algorithmic}
\usepackage{graphicx}
\usepackage{colortbl}
\usepackage{caption}
\usepackage{subcaption}
\usepackage{textcomp}
\usepackage{xcolor}
\usepackage{nicefrac}
\usepackage{bm}

\def\BibTeX{{\rm B\kern-.05em{\sc i\kern-.025em b}\kern-.08em
    T\kern-.1667em\lower.7ex\hbox{E}\kern-.125emX}}

\newcommand{\mbX}{{\mathbf{X}}}
\newcommand{\mbU}{{\mathbf{U}}}
\newcommand{\mbV}{{\mathbf{V}}}
\newcommand{\mbY}{{\mathbf{Y}}}
\newcommand{\mbhY}{{\hat{\mathbf{Y}}}}

\begin{document}

\title{Linear Progressive Coding for Semantic Communication using Deep Neural Networks}

\author{Eva Riherd, Raghu Mudumbai and Weiyu Xu
\thanks{The authors are with the Department of Electrical and Computer Engineering, University of Iowa, Iowa City, IA 52242, USA (e-mails: \{eva-riherd, raghuraman-mudumbai, weiyu-xu\}@uiowa.edu}
}

\maketitle

\begin{abstract}
We propose a general method for semantic representation of images and other 
data using progressive coding. Semantic coding allows for specific pieces of 
information to be selectively encoded into a set of measurements that can be highly 
compressed compared to the size of the original raw data. We consider a 
hierarchical method of coding where a partial amount of semantic information is 
first encoded a into a coarse representation of the data, which is then refined by 
additional encodings that add additional semantic information. Such hierarchical 
coding is especially well-suited for semantic communication i.e. transferring 
semantic information over noisy channels. Our proposed method can be considered as 
a generalization of both progressive image compression and source coding for 
semantic communication. We present results from experiments on the 
MNIST and CIFAR-10 datasets that show that progressive semantic coding can provide 
timely previews of semantic information with a small number of initial measurements 
while achieving overall accuracy and efficiency comparable to non-progressive methods.
\end{abstract}


\section{Introduction}

We consider the general problem of (linear) progressive semantic representation of data  
using deep neural networks for efficient data storage and communication. 

Semantic encoding means that we do not wish to store or transmit data in its raw form; instead, we wish to selectively encode certain meaningful information (``message") contained in the data. We consider the case where the message can be organized in a hierarchical sequence of categories. We seek to design a progressive encoding scheme where a coarse initial description of the message is augmented by refining descriptions. Given storage and communication constraints, our goal is to explore the tradeoffs between the amount of resources required for the initial coarse description and subsequent refinements.

\subsection{Related Work}

The idea of progressive coding has been most well-developed in the area of image processing. The concept of progressive image coding or compression was originally popularized \cite{chee1999survey} for efficiently transmitting images over slow Internet connections. Standards such as JPEG 2000 \cite{jpeg2000} allowed for encoding and transmitting images in a gradual manner, allowing for the display of lower-resolution versions while higher-resolution details are progressively transmitted.

This first generation of progressive image coding methods were primarily based on wavelet and frequency-domain representations \cite{image_coding_transform}. While some of this early works also attempted to take into account the human visual system to optimize the encoding \cite{human_visual_prog_coding}, the ability to minimize perceptual distortions \cite{perceptual} has been significantly enhanced by the more recent introduction of convolutional neural networks \cite{deepprog2}.

In fact, the new capabilities from neural networks have led to image coding methods that combine progressive encoding with {\it semantic data representation} \cite{semantic_image_coding2,DPCClassificationReconstruction}, wherein each step in progressive coding offers enhanced image by adding some
meaningful information that was previously missing. 
Simultaneously, the introduction of deep neural networks has also led to a renewed interest in semantic information processing with various types of data \cite{semantic_speech, semantic_text, semantic_image_coding1}. The idea is to use neural networks to selectively extract meaningful pieces of information from raw data for storage, processing and transmission. In communication engineering, this represents a major departure \cite{qin2022semantic} from the previously dominant Shannon model \cite{Shannonpaper} where semantics are ignored.

Recent work on {\it compressed learning} \cite{zisselman2018compressed} explores extracting semantic information from images using only a small number of measurements. It is well-known from the classical theory of compressed sensing that natural data such as images \cite{cs_images,Donoho:2006:1,candes_stable_2006} can be recovered from under-sampled measurements by taking advantage of sparsity. Compressed learning seeks to extract semantic information, rather than the image itself, from a minimal number of measurements \cite{cs_learning,optimalcompressionGao}.

\subsection{Contributions and Findings}

In this paper, we integrate the idea of progressive coding and compressed learning in semantic communications. Specifically, we used (linear, benefits of ``linear'' explained later) compressed measurements to efficiently encode semantic information in a progressive fashion for classification purpose at the receiver. Thus an initial small number of samples (measurements)  are used to encode information for coarse classifications, and later more samples are used to encode information for fine-grained classifications. Deep neural networks can be used to train projections for such measurements and to perform classifications using these compressed measurements.  

We report on a series of experiments on the MNIST and CIFAR-10 image datasets to illustrate this concept. In both experiments, we perform an initial coarse classification using smaller number of samples followed by a more detailed classification with more samples. Some key findings from these experiments are as follows.
\begin{enumerate}
\item We show that the raw signal data can be very significantly compressed into a small number of measurements to encode the semantic information of interest. This is consistent with the literature on semantic coding. Furthermore, the measurements involved only linear projections of the raw image data.

\item Our progressive classifiers are comparable in complexity (measured by number of layers and neurons) and achieves a similar performance (measured by classification accuracy) to non-progressive classifiers from the literature using the same number of measurements. Of course, our progressive classifiers are also able to provide a quick preview of a coarse classification.

\item There is a tradeoff between the accuracy of the initial coarse classification and the number of measurements used for the coarse encoding. The less obvious observation is that useful levels of accuracy can be achieved with a surprisingly small number of measurements. As an extreme case, for the MNIST dataset that we can make an initial prediction about an image label with $90 \%$ accuracy with {\it just one single linear measurement}.
\end{enumerate}

\section{Problem Statement}
Let $\mbX \in \mathbb{R}^N$ be a vector in a high-dimensional space such as a vectorized set of image pixels. Let $\mathbf{A}_k \in \mathbb{R}^{M_k\times N},~k=1 \dots K$, represent a sequence of measurement matrices that produce the sequence of measurements $\mbU_k \doteq \mathbf{A}_k \mbX$. The measurements $\mbU_k,~k=1 \dots K$, are transmitted over a noisy channel $P(\mbV_k | \mbU_k)$ and the resulting noisy measurements $\mbV_k,~k=1 \dots K$ are processed by machine-learning or other prediction algorithms to produce a sequence of predicted labels $\mbhY_1 \doteq g_1(\mbV_1),~\mbhY_2 \doteq g_2(\mbV_1,\mbV_2),~ \dots,~ \mbhY_k \doteq g_k(\mbV_1,\mbV_2,\dots,\mbV_k),~\dots,~\mbhY_K \doteq g_K(\mbV_1,\mbV_2,\dots,\mbV_K)$.

The true labels $\mbY_k = f_k(\mbX),~k= 1 \dots K$ represent a sequence of refinements of semantic information contained in $\mbX$, where $\mbY_1$ and $\mbY_K$ represents a very coarse-grained and fine-grained label respectively. Our goal is to eventually recover the fine-grained label $\mbY_K$. However, we would also like to obtain quick previews and successive refinements in the form of the coarse-grained labels $\mbY_1,~\mbY_2, \dots$ similar to how progressive image coding gradually generates a high resolution image by refining an initial low resolution image.

In general, we aim to have accurate prediction of $\mbY_k$, $k=1 \dots K$. We prioritize having timely classifications for $\mbY_k$'s with lower index $k$ with earliest-received batch of samples at the communication receiver. The utility of the communication receiver can be modeled by a weighted sum of mutual information:
\begin{align}
\sum_{k=1}^{K} \lambda_k I(\mbV_1, \mbV_2,~ \dots, \mbV_k ; \mbY_{k}), 
\label{eq:objective}
\end{align}
where $\lambda_k$'s are adjustable non-negative parameters putting different priorities on the different grain-level tasks. For example, if $K=2$, $\lambda_1\gg \lambda_2>0$ implies that the first set of measurements $\mbV_1$ needs to give highest accuracy for decoding label $\mbY_1$; moreover, conditioned on that, the 2nd batch of measurements are required to give highest accuracy for decoding label $\mbY_2$, when combined with the existing 1st batch of measurements.

Our goal is to design a progressive encoding (sampling) scheme that optimize \eqref{eq:objective}. {Note that generally we can use   
non-linear projections (for example, projections through neural network) of $\mbX$ to obtain these compressed projection $\mbU_k$'s. However, besides linear measurements being simple to implement in low-power sensors or devices, we particularly propose linear projections for the following reasons. Firstly, when the total number of (noiseless) samples $\sum_{k=1}^{K} M_k=N$, one can simply use matrix inverse to fully recover the full data $\mbX$; however, for general non-linear measurements, we do not have efficient algorithms that theoretically guarantee fully recovering $\mbX$. Secondly, linear measurements can be more robust against adversarial attacks when compared with non-linear measurements obtained through neural networks. Even when the number of linear samples $\sum_{k=1}^{K} M_k \ll N$, one can still use sparsity-based compressed sensing to recover the full signal with (adversarial) robustness guarantees.      
}

\section{General architecture of (Linear) progressive semantic coding}

\begin{figure}
		\includegraphics[width=0.5\textwidth, height=0.4\textwidth]{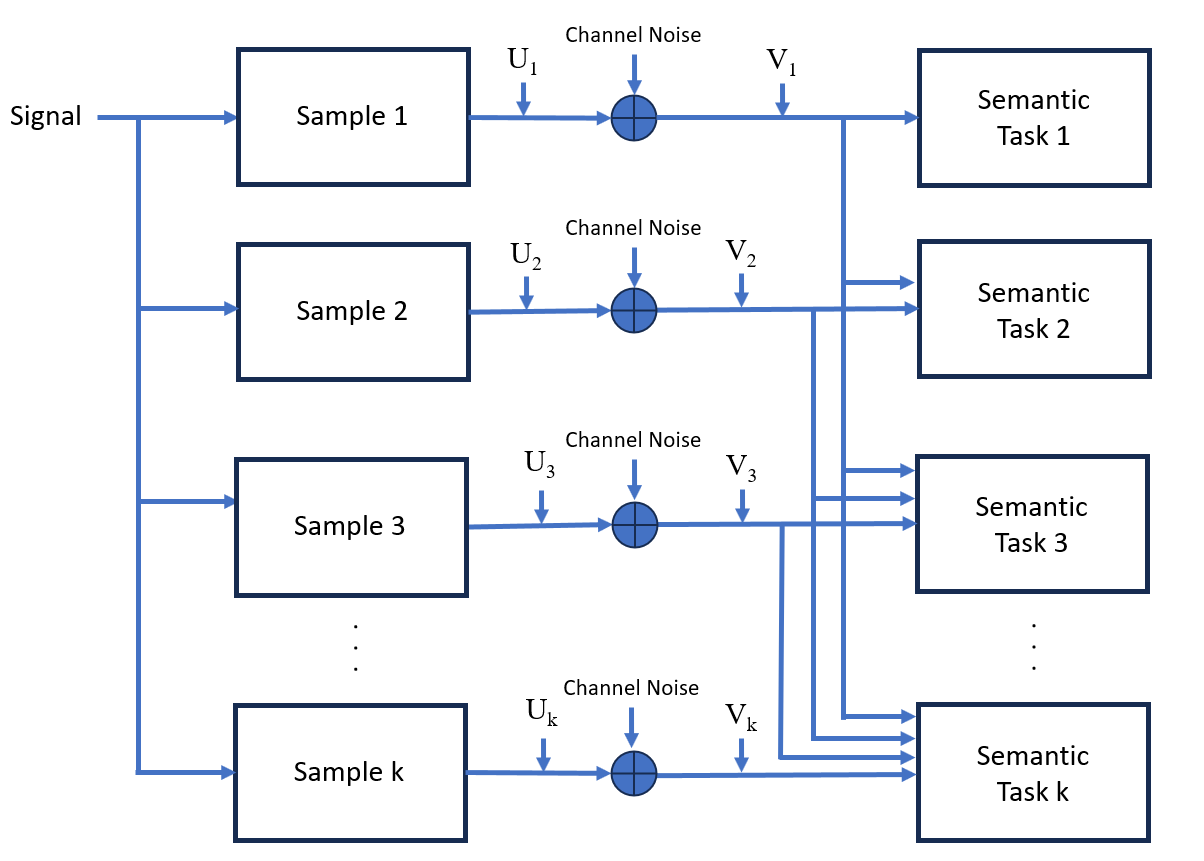}
		\caption{Illustration of progressive coding for semantic communication}
		\label{fig:progcoding}
\end{figure}

Figure \ref{fig:progcoding} describes a general architecture for linear progressive coding for semantic communication. In this architecture, the first batch of linear samples $\mbU_1$ is optimized (trained) to give best performance for coarser-level Semantic Task 1. Now with $\mbU_1$ fixed, we train another batch of linear samples $\mbU_2$ such that when combined with $\mbU_1$, we have the best performance for finer-level Semantic Task 2. Note that we cannot re-optimize the first batch of samples $\mbU_1$ for Semantic Task 2, because these $\mbU_1$ samples are already optimized for Task 1 and then fixed. This can extend to more levels of tasks. 

\section{Numerical Results}

We designed and performed a series of experiments with the MNIST and CIFAR-10 datasets to demonstrate our idea of progressive semantic coding.

\subsection{Experiment Setup}

\begin{figure}
		\includegraphics[width=0.5\textwidth, height=0.4\textwidth]{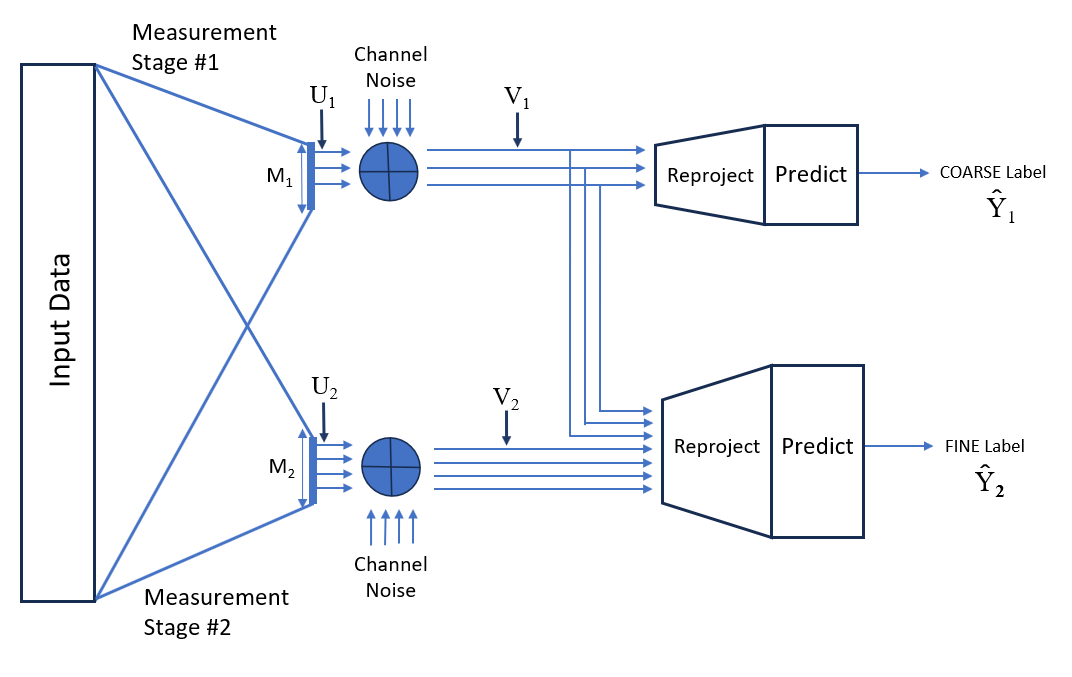}
		\caption{ MNIST Neural Network Architecture }
		\label{fig:nnarch}
\end{figure}

Figure \ref{fig:nnarch} shows the experiment design in block diagram. A transmitter takes a high-dimensional input signal (e.g. an MINST or CIFAR-10 image) and performs a small number $M_1$ of linear measurements on the input signal. These measurements are sent over a noisy communication link to a receiver which feeds the noisy measurements to a neural network classifier to produce a coarse initial classification.

The transmitter then performs an additional number of $M_2$ linear measurements on the input signal which are also then sent to the receiver over the noisy link. The receiver feeds all $M_1+M_2$ noisy measurements into a second neural network classifier that produces a final fine-grained classification that represents a refinement of the initial prediction.

\subsection{Noise-free Experiments with the MNIST dataset}

The MNIST dataset is a widely used collection of $28 \times 28$ pixel grayscale images of handwritten digits (0-9) designed for training and evaluating machine learning models for digit recognition. For our progressive coding experiment, we split up the digit recognition task into a 2-step process: first we perform a coarse prediction of whether the digit in the image is even or odd, and in the second step, refine the initial coarse even/odd prediction into a full 0-9 digit prediction.

We now describe the training process used for the experiment. We first trained an end-to-end neural network for the coarse prediction. The weights of the linear encoder that produces $M_1$ linear measurements as well as the reprojection and prediction layers that produce the coarse prediction are optimized using stochastic gradient descent during this training. For the fine prediction, we perform another round of training, where the weights for the initial $M_1$ measurements are kept fixed, while the weights for the second set of $M_2$ measurements as well as the reprojection and prediction layers of the second neural network are optimized for the full digit recognition.

\begin{table*}[h]
\centering
\begin{tabular}{|| c c || c c c c | >{\columncolor[gray]{0.8}} c >{\columncolor[gray]{0.8}} c | c c c c c||} 
 \hline
 a & b & Col 1 & Col 2 & Col 3 & Col 4 & Col 5 & Col 6 & Col 7 & Col 8 & Col 9 & Col 10 & Col 11\\ [0.5ex] 
 \hline\hline
 1 & 8 & 0.9012 & 0.3017 & 0.7048 & 0.4983 & 0.9039 & 0.9550 & 0.9683 & 0.8265 & 0.9770 & 0.9594 & 0.9733 \\ 
 2 & 8 & 0.9435 & 0.5391 & 0.8634 & 0.7237 & 0.9320 & 0.9587 & 0.9772 & 0.8988 & 0.9776 & 0.9611 & 0.9751 \\
 3 & 8 & 0.9662 & 0.6335 & 0.9033 & 0.8372 & 0.9493 & 0.9663 & 0.9825 & 0.9095 & 0.9811 & 0.9631 & 0.9796 \\
 4 & 8 & 0.9692 & 0.6939 & 0.9480 & 0.8879 & 0.9649 & 0.9631 & 0.9829 & 0.9390 & 0.9810 & 0.9647 & 0.9820 \\
5 & 8 & 0.9689 & 0.7060 & 0.9605 & 0.9196 & 0.9717 & 0.9651 & 0.9782 & 0.9344 & 0.9828 & 0.9697 & 0.9840 \\ [1ex] 
10 & 8 & 0.9787 & 0.8975 & 0.9769 & 0.9616 & 0.9819 & 0.9728 & 0.9832 & 0.9584 & 0.9863 & 0.9714 & 0.9871 \\ [1ex] 
20 & 8 & 0.9842 & 0.9591 & 0.9867 & 0.9726 & 0.9863 & 0.9735 & 0.9868 & 0.9675 & 0.9876 & 0.9758 & 0.9868 \\ [1ex] 
5 & 5 & 0.9715 & 0.7866 & 0.9600 & 0.9223 & 0.9747 & 0.9577 & 0.9809 & 0.9024 & 0.9814 & 0.9624 & 0.9791 \\ [1ex] 
\rowcolor{lightgray}
5 & 10 & 0.9716 & 0.7396 & 0.9600 & 0.9225 & 0.9684 & 0.9683 & 0.9833 & 0.9463 & 0.9842 & 0.9727 & 0.9853 \\ [1ex] 
5 & 20 & 0.9703 & 0.7444 & 0.9583 & 0.9237 & 0.9684 & 0.9778 & 0.9863 & 0.9664 & 0.9879 & 0.9756 & 0.9883 \\ [1ex] 
5 & 30 & 0.9719 & 0.7128 & 0.9551 & 0.9187 & 0.9702 & 0.9747 & 0.9884 & 0.9724 & 0.9878 & 0.9781 & 0.9884 \\ [1ex] 
5 & 39 & 0.9702 & 0.7365 & 0.9582 & 0.9211 & 0.9621 & 0.9778 & 0.9859 & 0.9730 & 0.9883 & 0.9745 & 0.9867 \\ [1ex] 
1 & 1 & 0.9034 & 0.3038 & 0.7474 & 0.4842 & 0.9041 & 0.6375 & 0.9459 & 0.4811 & 0.8869 & 0.7457 & 0.9044 \\ [1ex] 
1 & 2 & 0.9038 & 0.3060 & 0.6942 & 0.4951 & 0.9045 & 0.8070 & 0.9627 & 0.6177 & 0.9106 & 0.8325 & 0.9163 \\ [1ex] 
1 & 3 & 0.9037 & 0.3091 & 0.7666 & 0.4852 & 0.9018 & 0.8744 & 0.9567 & 0.7270 & 0.9389 & 0.8949 & 0.9353 \\ [1ex] 
1 & 4 & 0.9028 & 0.3112 & 0.7561 & 0.4740 & 0.9040 & 0.9125 & 0.9721 & 0.7404 & 0.9584 & 0.9174 & 0.9526 \\ [1ex] 
1 & 5 & 0.9022 & 0.3090 & 0.7527 & 0.4787 & 0.9038 & 0.9293 & 0.9685 & 0.8200 & 0.9684 & 0.9363 & 0.9598 \\ [1ex] 
1 & 10 & 0.9036 & 0.3045 & 0.7395 & 0.5109 & 0.9036 & 0.9616 & 0.9806 & 0.9013 & 0.9816 & 0.9617 & 0.9801 \\ [1ex] 
1 & 20 & 0.9043 & 0.3031 & 0.7703 & 0.5046 & 0.9012 & 0.9725 & 0.9867 & 0.9622 & 0.9876 & 0.9698 & 0.9865 \\ [1ex] 
1 & 30 & 0.8982 & 0.2993 & 0.6994 & 0.5057 & 0.9010 & 0.9750 & 0.9863 & 0.9702 & 0.9875 & 0.9783 & 0.9867 \\ [1ex] 
1 & 40 & 0.9025 & 0.3144 & 0.7642 & 0.4527 & 0.9031 & 0.9761 & 0.9871 & 0.9700 & 0.9879 & 0.9773 & 0.9883 \\ [1ex] 
1 & 50 & 0.9047 & 0.3062 & 0.7474 & 0.4795 & 0.9032 & 0.9759 & 0.9867 & 0.9705 & 0.9900 & 0.9775 & 0.9884 \\ [1ex] 
 \hline
\end{tabular}
\caption{MNIST Results.}
\label{table:mnist}
\end{table*}

We performed an extensive set of experiments for various different values of $M_1=a,M_2=b$ under noise-free conditions i.e. the inputs $\mbV_k$ to the decoder are identical to the measurements $\mbU_k$. The results are shown in Table \ref{table:mnist}. A brief description of the entries in Table \ref{table:mnist} follows. Each row of Table \ref{table:mnist} has $11$ columns of accuracy numbers for a sequence of experiments for a common set of $(a,b)$ parameter values. Col 1 shows the accuracy of the coarse (even/odd) prediction with $M_1=a$ measurements all optimized for coarse prediction, and Col 2 shows the fine ($0-9$) digit prediction accuracy using the same $M_1=a$ measurements as Col 1. Col 3 and Col 4 show coarse and fine prediction accuracy respectively with $M_1=a$ measurements all optimized for fine prediction.

Col 5 shows coarse prediction accuracy with $M_1=a$ measurements optimized for coarse prediction, and Col 6 shows fine prediction accuracy with $M_2=b$ additional measurements optimized for fine prediction.

Col 7 shows the coarse prediction accuracy using $M_1=a+b$ measurements all optimized for coarse prediction, and Col 8 shows the accuracy of fine predictions based on the same measurements as Col 7. Col 9 and Col 10 show the accuracy of coarse and fine predictions using $M_2=a+b$ measurements all optimized for fine prediction.

{\bf Discussion.} The coarse and fine prediction accuracy numbers reported in Columns 5, 6 respectively of Table \ref{table:mnist} represent the performance of our proposed progressive coding method under noise-free conditions. The remaining columns provide various benchmarks for comparison. Remarkably, it is possible to achieve $90 \%$ accuracy for even/odd prediction based on just one linear measurement i.e. a neural network decoder is able to predict whether the digit in the image is even or odd with $90 \%$ accuracy using just one well-chosen linear projection of the pixels of the image! Column 7 serves as an upper-bound for the accuracy of the coarse prediction using a total of $M_1+M_2 = a+b$ measurements, and likewise Column 10 serves as an upper-bound for fine prediction using $M_1+M_2 = a+b$ total measurements.

Consider the row corresponding to $a=5,~b=10$ in Table \ref{table:mnist}. Col 1 shows that the $M_1=a=5$ initial measurements achieve even/odd prediction accuracy of more than $97 \%$. However, from Col 2, we see that these initial $5$ measurements, being optimized for even/odd prediction, can only achieve a $74 \%$ accuracy for $0-9$ digit prediction. This number improves very substantially to almost $97 \%$ with the addition of $M_2=b=10$ additional measurements as seen from Col 6. This overall accuracy is based on a total of $M_1+M_2=a+b=15$ measurements of which $5$ are optimized for the initial coarse prediction task. If we optimize all $15$ measurements for the fine $0-9$ digit prediction task, the accuracy improves only slightly as seen from Col 10.

The difference between Col 6 and Col 10 can be thought of as a penalty for the progressive coding: the slightly lower accuracy in Col 6 is the price we pay for being able to make a quick even/odd prediction. We can see that this penalty is consistently small.

\subsection{Effect of Channel Noise}
The results in Table \ref{table:mnist} were from experiments under noise-free conditions which are of course not realistic for a communication setting. In general, the neural network classifier does not have access to the linear measurements $\mbU_k$ directly, but only to noise corrupted copies $\mbV_k$ of these measurements.

To study the effect of noise, we modified the noise-free experiments by retraining the classifiers with noisy measurements. Specifically, for a fixed noise level $\sigma_w^2$, we added several random realizations of white Gaussian noise to the measurements from each training image: $\mbV_k \equiv \mbU_k + \mathbf{W}_k,~\mathbf{W}_k \sim N(\mathbf{0}, \sigma_w^2 \mathbb{I}_{M_k})$. We then retrained the weights for the reprojection and prediction layers for the coarse and fine prediction networks, and then tested the accuracy of the newly trained networks with noisy measurements on test images. This process was repeated for several different noise levels $\sigma_w^2$. Note that the linear measurements were not modified by this training process. In particular, we use the same linear measurements as the noise-free experiments for new experiments with noise.

\begin{figure}
\includegraphics[width=0.45\textwidth, height=0.25\textwidth]{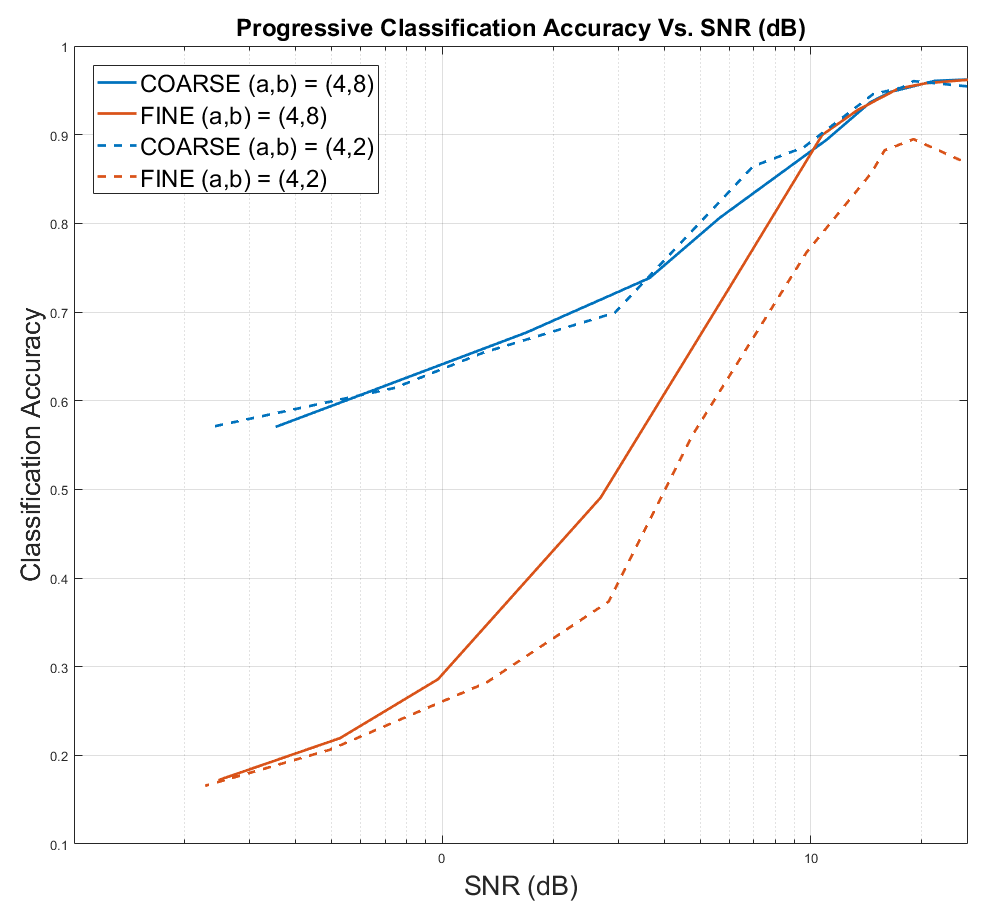}
\caption{Coarse and Fine Classification Accuracy vs SNR.}
		\label{fig:snr1}
\end{figure}

Figure \ref{fig:snr1} shows the accuracy of the coarse and fine prediction as a function of SNR. As expected, the accuracy improves with SNR and essentially matches the performance in the noise-free case for SNRs above $13$ dB or so.

\subsection{CIFAR-10 Results}
 For the CIFAR-10 datasets, 
 we focus on classification of 4 classes: deer, horse, automobile and truck.  The coarse classification is to classify whether it is an animal or vehicle. The fine classification is to further distinguish whether it is a deer or horse; an automobile or truck. We adopt the same progressive architecture for coarse and fine classifications. If we use $M_1=102$ and $M_2=102$, we have $87.6\%$ accuracy for coarse classification using the first $M_1$ measurements optimized for coarse classification, and $71.1\%$ accuracy for fine classification using the $M_1+M_2$ measurements. However, if $M_2=922$, we can increase the accuracies respectively to  $96.9 \%$ and $92.6 \%$, if we use these $M_1+M_2$ measurements for coarse and fine classification. Using fewer samples, one can already achieve decent accuracy for quicker coarse classification.

\clearpage
\bibliographystyle{IEEEtran}
\bibliography{refs.bib, CS_ref.bib}

\begin{thebibliography}{10}
\providecommand{\url}[1]{#1}
\csname url@samestyle\endcsname
\providecommand{\newblock}{\relax}
\providecommand{\bibinfo}[2]{#2}
\providecommand{\BIBentrySTDinterwordspacing}{\spaceskip=0pt\relax}
\providecommand{\BIBentryALTinterwordstretchfactor}{4}
\providecommand{\BIBentryALTinterwordspacing}{\spaceskip=\fontdimen2\font plus
\BIBentryALTinterwordstretchfactor\fontdimen3\font minus
  \fontdimen4\font\relax}
\providecommand{\BIBforeignlanguage}[2]{{%
\expandafter\ifx\csname l@#1\endcsname\relax
\typeout{** WARNING: IEEEtran.bst: No hyphenation pattern has been}%
\typeout{** loaded for the language `#1'. Using the pattern for}%
\typeout{** the default language instead.}%
\else
\language=\csname l@#1\endcsname
\fi
#2}}
\providecommand{\BIBdecl}{\relax}
\BIBdecl

\bibitem{chee1999survey}
Y.-K. Chee, ``Survey of progressive image transmission methods,''
  \emph{International journal of imaging systems and technology}, vol.~10,
  no.~1, pp. 3--19, 1999.

\bibitem{jpeg2000}
M.~Marcellin, M.~Gormish, A.~Bilgin, and M.~Boliek, ``An overview of
  jpeg-2000,'' in \emph{Proceedings DCC 2000. Data Compression Conference},
  2000, pp. 523--541.

\bibitem{image_coding_transform}
H.~Malvar, ``Fast progressive image coding without wavelets,'' in
  \emph{Proceedings DCC 2000. Data Compression Conference}, 2000, pp. 243--252.

\bibitem{human_visual_prog_coding}
B.~Chitprasert and K.~Rao, ``Human visual weighted progressive image
  transmission,'' \emph{IEEE Transactions on Communications}, vol.~38, no.~7,
  pp. 1040--1044, 1990.

\bibitem{perceptual}
Z.~Wang, Q.~Li, and X.~Shang, ``Perceptual image coding based on a maximum of
  minimal structural similarity criterion,'' in \emph{IEEE International
  Conference on Image Processing}, vol.~2, 2007, pp. II -- 121--II -- 124.

\bibitem{deepprog2}
C.~Cai, L.~Chen, X.~Zhang, G.~Lu, and Z.~Gao, ``A novel deep progressive image
  compression framework,'' in \emph{2019 Picture Coding Symposium (PCS)}, 2019,
  pp. 1--5.

\bibitem{semantic_image_coding2}
D.~Huang, X.~Tao, F.~Gao, and J.~Lu, ``Deep learning-based image semantic
  coding for semantic communications,'' in \emph{2021 IEEE Global
  Communications Conference (GLOBECOM)}, 2021, pp. 1--6.

\bibitem{DPCClassificationReconstruction}
Z.~Lei, P.~Duan, X.~Hong, J.~F.~C. Mota, J.~Shi, and C.-X. Wang, ``Progressive
  deep image compression for hybrid contexts of image classification and
  reconstruction,'' \emph{IEEE Journal on Selected Areas in Communications},
  vol.~41, no.~1, pp. 72--89, 2023.

\bibitem{semantic_speech}
Z.~Weng, Z.~Qin, and G.~Y. Li, ``Semantic communications for speech signals,''
  in \emph{ICC 2021 - IEEE International Conference on Communications}, 2021,
  pp. 1--6.

\bibitem{semantic_text}
N.~Farsad, M.~Rao, and A.~Goldsmith, ``Deep learning for joint source-channel
  coding of text,'' in \emph{2018 IEEE International Conference on Acoustics,
  Speech and Signal Processing (ICASSP)}, 2018, pp. 2326--2330.

\bibitem{semantic_image_coding1}
D.~Huang, F.~Gao, X.~Tao, Q.~Du, and J.~Lu, ``Toward semantic communications:
  Deep learning-based image semantic coding,'' \emph{IEEE Journal on Selected
  Areas in Communications}, vol.~41, no.~1, pp. 55--71, 2023.

\bibitem{qin2022semantic}
Z.~Qin, X.~Tao, J.~Lu, W.~Tong, and G.~Y. Li, ``Semantic communications:
  Principles and challenges,'' 2022.

\bibitem{Shannonpaper}
\BIBentryALTinterwordspacing
C.~E. Shannon, ``A mathematical theory of communication,'' \emph{SIGMOBILE Mob.
  Comput. Commun. Rev.}, vol.~5, no.~1, p. 3–55, jan 2001. [Online].
  Available: \url{https://doi.org/10.1145/584091.584093}
\BIBentrySTDinterwordspacing

\bibitem{zisselman2018compressed}
E.~Zisselman, A.~Adler, and M.~Elad, ``Compressed learning for image
  classification: A deep neural network approach,'' in \emph{Handbook of
  Numerical Analysis}.\hskip 1em plus 0.5em minus 0.4em\relax Elsevier, 2018,
  vol.~19, pp. 3--17.

\bibitem{cs_images}
L.~Gan, ``Block compressed sensing of natural images,'' in \emph{2007 15th
  International Conference on Digital Signal Processing}, 2007, pp. 403--406.

\bibitem{Donoho:2006:1}
D.~Donoho, ``Compressed sensing,'' \emph{IEEE Trans. Inf. Theory}, vol.~52,
  no.~4, pp. 1289--1306, Apr. 2006.

\bibitem{candes_stable_2006}
\BIBentryALTinterwordspacing
E.~Cand{\`e}s, J.~Romberg, and T.~Tao, ``\BIBforeignlanguage{en}{Stable signal
  recovery from incomplete and inaccurate measurements},''
  \emph{\BIBforeignlanguage{en}{Comm. Pure Appl. Math.}}, vol.~59, no.~8, pp.
  1207--1223, Aug. 2006. [Online]. Available:
  \url{http://onlinelibrary.wiley.com/doi/10.1002/cpa.20124/abstract}
\BIBentrySTDinterwordspacing

\bibitem{cs_learning}
R.~Calderbank and S.~Jafarpour, ``Finding needles in compressed haystacks,'' in
  \emph{2012 IEEE International Conference on Acoustics, Speech and Signal
  Processing (ICASSP)}, 2012, pp. 3441--3444.

\bibitem{optimalcompressionGao}
J.~Gao, A.~Tang, and W.~Xu, ``Optimal compression for minimizing classification
  error probability: An information-theoretic approach,'' in \emph{ICASSP 2023
  - 2023 IEEE International Conference on Acoustics, Speech and Signal
  Processing (ICASSP)}, 2023, pp. 1--5.

\end{thebibliography}

\end{document}